# Machine learning the quantum flux-flux correlation function for catalytic surface reactions

Brenden G. Pelkie[1*] and Stéphanie Valleau,[1*]

[1]*Department of Chemical Engineering, University of Washington, Seattle, Washington 98195, United States*
*corresponding authors: bgpelkie@uw.edu, valleau@uw.edu

**ABSTRACT**

A dataset of fully quantum flux-flux correlation functions and reaction rate constants was constructed for organic heterogeneous catalytic surface reactions. Gaussian process regressors were successfully fitted to training data to predict previously unseen test set reaction rate constant products and Cauchy fits of the flux-flux correlation function. The optimal regressor prediction mean absolute percent errors were on the order of 0.5% for test set reaction rate constant products and 1.0% for test set flux-flux correlation functions. The Gaussian process regressors were accurate both when looking at kinetics at new temperatures and reactivity of previously unseen reactions and provide a significant speedup respect to the computationally demanding time propagation of the flux-flux correlation function.

**INTRODUCTION**

Many theories have been developed to approximate quantum reaction rate constants or lower the cost of their computation.[1,2] Indeed, the curse of dimensionality has impeded the calculation of quantum reaction rate constants dynamically: the cost scales exponentially with the degrees of freedom.[3,4] To date, the largest fully quantum scattering calculations can only account for systems of approximately six atoms.[5] In catalysis light-weight atoms such as hydrogen are often diffusing on metal surfaces. Hence quantum effects such as tunneling must be accounted for when computing reaction rate constants. Optimal catalysts may enable a reduction of industrially produced toxic byproducts,[6] the removal of carbon dioxide[7] from the atmosphere and so forth. By understanding the type of reactant surface interactions[8], the reaction mechanisms,[9] and the corresponding reaction rate constants, we can establish which factors lead to optimal reactivity.

In recent years machine learning (ML) has successfully been employed to accelerate the evaluation of a variety of chemical and molecular properties.[10] For kinetics, the main bottleneck has been the lack of large representative datasets of reaction rate constants. These are necessary to train machine learning algorithms to predict the reaction rate constant. Recently some kinetic datasets have been generated for non-catalytic systems,[11–14] and contain activation energies,[13] as well as quantum reaction rates for one dimensional systems.[12] For catalytic surface reactions, Catalysis-Hub[15] has become a great resource, yet activation energies and minimum energy paths are only available for few reactions. Nonetheless with these limited datasets, supervised machine learning algorithms have successfully been used[16] to predict reactant and transition state partition functions,[17,18] Gibbs free activation energies,[19] activation energies,[20] and quantum reaction rate constants[12,21] for small systems. They have also been employed to accelerate the search for minimum energy paths.[22–25] For catalytic systems, machine learning has been used in a variety of contexts[26–28] such as predicting adsorption energies,[29] but little has been carried out for reaction rate constants, due to the lack of training data.

In this work we generated a small set of exact quantum flux-flux correlation reaction rate constants[30] for heterogenous catalytic surface reactions. With this dataset we trained gaussian process regressors (GPRs) to predict the reaction rate constant and fits to the dynamic flux-flux correlation function. The predicted fits were also used to compute the reaction rate constant. We will describe our workflow and discuss the results in the following subsections.

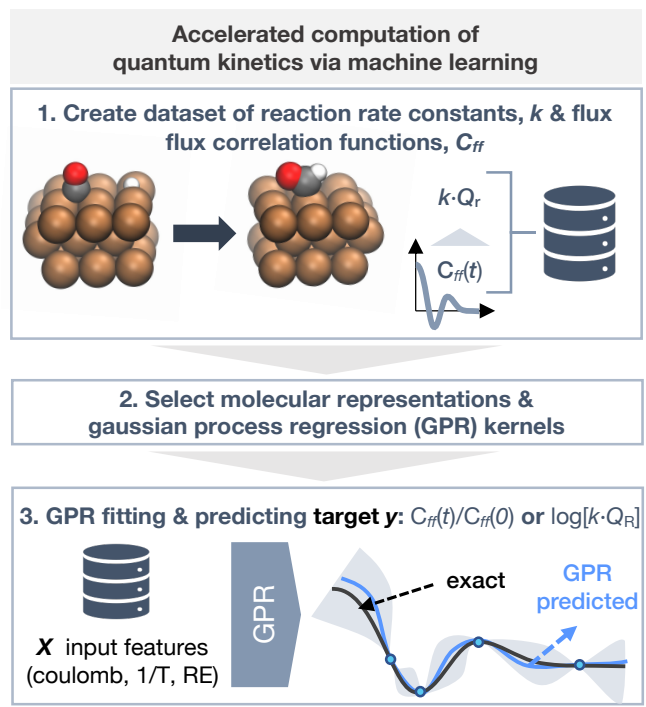

Figure 1: Flow chart of workflow. Panel 1. A dataset of reaction rate constants and flux-flux correlation functions was generated. Panel 2. With the dataset several gaussian process regressors, GPR, were trained using a series of input features and kernels to identify the optimal set. Panel 3. With the optimal kernel and input features, $X$, previously unseen test set reaction rate constant products with the reactant partition function, $\log(kQ_R)$, and parameters for fits to the scaled flux-flux correlation functions, $C_{ff}(t)/C_{ff}(0)$ were predicted. In panel 3, RE stands for reaction energy i.e. the difference between product and reactant zero point energies.

**RESULTS AND DISCUSSION**

**Dataset of quantum flux-flux correlation functions and reaction rate constants**

A set of fourteen gas phase heterogeneous catalytic surface reactions was selected from Catalysis-Hub (Table 1).[15] For these reactions, either a transition state geometry or a minimum energy path (MEP) were available, together with reactant and product geometries and energies. For reactions with missing MEPs, Eckart or skew normal functions were fitted to the energies of the reactants, products and transition states. The small size of the dataset was necessary due to the computational time required for the evaluation of the quantum reaction rate constants. Computing the single flux-flux correlation reaction rate constant for 55 one dimensional reactive pathways took six months of work. This calculation would take several months for a single path when expanding to two or three dimensional potential energy surfaces.

With the potential energy paths we defined the Hamiltonians, $\widehat{H}$, Heaviside dividing surface operators, $\widehat{h}$, flux operators, $\widehat{F} = {}^i/_\hbar [\widehat{H}, \widehat{h}]$, time evolution operators, $\widehat{U}$ and Boltzmann operators $e^{\beta \widehat{H}}$ in the canonical ensemble for each system using the sinc basis set discrete variable representation (DVR).[31,32]

In this representation, operators are defined by matrices with elements evaluated using a basis set of sinc functions, $\text{sinc}(x) = \sin(x)/x$. Each function is centred at one point on a uniformly spaced set of $N_{DVR}$ grid points $x_j = x_{min} + \Delta x(j-1)$. The grids were centred such that the maximum of the potential energy path was in position $x = 0$ and the edges of the grid at positions $\pm L$. The products of the quantum reaction rate constant $k(T)$ with the reactant partition function $Q_r(T)$ were then computed (eq 1) by integrating the quantum flux-flux correlation function $C_{ff}(t;T)$ numerically (Eq 1)[30]

$$k(T) \cdot Q_R(T) = \int_0^{+\infty} C_{ff}(t) dt \quad (1)$$

$$C_{ff}(t;T) = \Re\left\{ Tr\left[ e^{-\beta \hat{H}} \cdot \hat{F} \cdot e^{i\hat{H}t/\hbar} \cdot \hat{F} \cdot e^{-i\hat{H}t/\hbar} \right] \right\}. \quad (2)$$

The upper bound of the integral in Eq. 1 was determined by finding the time right before the wavefunction reaches the edges of the grid, for more information see Supporting Information. The cost of the calculations is determined by the size of the grid $N_{DVR}$. Computing $C_{ff}(t;T)$ using Eq. 2 is exact at any given time and does not require previous values in time. However, it requires the computation of the time evolution operator at each time step as well as the products with the flux and Boltzmann operators.

Table 1: List of the reactions present in the dataset. Reactions were taken from Catalysis-Hub.

| $N_{react}$ | Reaction | Catalyst | Surface | $E_a$ [kcal / mol] |
|---|---|---|---|---|
| 1 | $CH^* + ^* \rightarrow C^* + H^*$ | Rh | 111 | 33.53 |
| 2 | $COH^* + ^* \rightarrow C^* + OH^*$ | Rh | 111 | 27.93 |
| 3 | $CHOH^* + ^* \rightarrow CHO^* + H^*$ | Rh | 111 | 19.17 |
| 4 | $CH_3^* + ^* \rightarrow CH_2^* + H^*$ | Rh | 211 | 10.25 |
| 5 | $CH_2OH^* + ^* \rightarrow CHOH^* + H^*$ | Pt | 111 | 27.00 |
| 6 | $CH_2^* + ^* \rightarrow CH^* + H^*$ | Ir | 111 | 2.55 |
| 7 | $CH_3^* + ^* \rightarrow CH_2^* + H^*$ | Ir | 111 | 14.17 |
| 8 | $CH_3^* + ^* \rightarrow CH_2^* + H^*$ | Pt | 111 | 24.73 |
| 9 | $CHOH^* + ^* \rightarrow HCO^* + H^*$ | Ag | 111 | 14.67 |
| 10 | $CHOH^* + ^* \rightarrow CH^* + OH^*$ | Ir | 111 | 12.29 |
| 11 | $CHO^* + ^* \rightarrow CO^* + H^*$ | Pd | 111 | 2.85 |
| 12 | $COH^* + H^* \rightarrow CHOH^*$ | Cu | 100 | 17.01 |
| 13 | $CO^* + H^* \rightarrow CHO^*$ | Cu | 100 | 22.36 |
| 14 | $CH^* + H^* \rightarrow CH_2^*$ | Cu | 100 | 17.01 |

It is worth noting that the boundaries of the grid can lead to non-physical reflection in the dynamics of the flux-flux correlation function. To identify these and find the optimal grid parameters, calculations were carried out at fixed grid spacing and increasing grid width. We also searched for and identified the minimum temperature for which a convergent $C_{ff}(t;T)$ function could be computed for each reaction. Details of the DVR parameters used for each temperature and reaction are provided in the supporting information.

For each reaction, four temperatures were randomly selected in the range [150-400] K. The entire dataset includes 14 fitted reaction minimum energy paths, 55 reaction rate constant products, 55 flux-flux correlation functions as a function of time, reaction energies and activation energies. It can be downloaded from Zenodo.[33]

**Gaussian process regression of reaction rate constant and fit parameter for the flux flux correlation function**

Given the small size of our dataset we chose to use Gaussian process regression (GPR), a Bayesian non-parametric ML model, as our machine learning model (Figure 2).[16] With GPRs, given some input data, $X$ (training data set input features) a posterior distribution over functions, $f(x)$ can be inferred. With this posterior one can make predictions, $y^*$ (targets) from new inputs, $X^*$ (test data set input features).

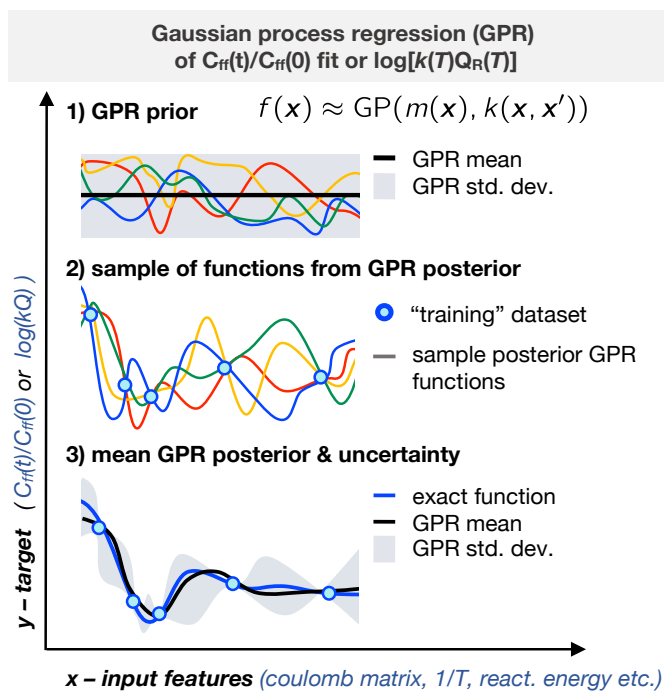

Figure 2: Panel 1: Gaussian process regressor prior distribution with zero mean and a kernel of choice. Panel 2: Sample functions from the posterior, generated from the prior (Panel 1) based on the training data. Panel 3: Mean of the trained GPR (black line) and prediction standard deviation (shaded area) respect to exact target function (blue line).

Here our predictions or "targets" are the natural logarithm of the product of the reaction rate constant with the canonical reactant partition function $k(T) \cdot Q_R(T)$ or a scale parameter for Cauchy fit to the ratio $C_{ff}(t;T)/C_{ff}(0;T)$. We will discuss this second scale parameter target in more detail in the next subsection. The gaussian process is defined as the prior for the regression function; it depends on a mean function and kernel or covariance function (Figure 2, panel 1). The joint distribution of the training and test outputs is defined by using the prior. Then we can obtain the posterior distribution by conditioning the joint gaussian prior distribution on the observations, $X, y$ (Figure 2, panels 2-3). The gaussian process's prediction ability depends on the choice of kernel and estimation of optimal kernel parameters. In the next subsections we will discuss how the train and test datasets were obtained and how the kernel and kernel parameters were identified.

**Train and test datasets**

Two train and test datasets were generated from the full dataset. To investigate the ability of Gaussian process regressors to predict kinetic quantities as a function of temperature (temperature-wise train test split) and for unseen reactions (reaction-wise train test split).

In the first case the data was split by temperature to investigate whether a GPR could learn the quantum reaction rate constant or flux-flux correlation function targets at new temperatures. For each reaction, a row from the dataset was selected for the test set by randomly choosing one temperature. The rest of the dataset rows for that reaction were placed in the train set.

In the second case the data was split by reaction to analyse whether a trained GPR could predict the quantum reaction rate constant or flux-flux correlation function targets for an entirely new reaction. Here 4 reactions were chosen to be part of the test set and 10 of the train set. Specifically, reactions 2, 7, 8 and 12 were placed in the test set. Due to the small size of the overall dataset, this choice was done by hand to ensure that every metal catalyst present in the test set was represented in the training set. For silver and palladium (reactions 9 and 11) only one example was present. These were therefore added to the training set.

As discussed earlier the target was either the natural logarithm of the product of the quantum reaction rate constant with the reactant canonical partition function, $k(T) \cdot Q_R(T)$, or the scale parameter $s$ for a Cauchy fit (Eq 3) of the ratio $C_{ff}(t;T)/C_{ff}(t;0)$. We had initially tried predicting exact $C_{ff}(t;T)$ values but the error was very large. Hence, we opted to scale $C_{ff}(t;T)$ by its value at time zero and to fit it to a Cauchy distribution function:

$$C_{ff}(t;T) \approx C_{ff}(0;T) \frac{1}{\pi s \left(1 + \left(\frac{t - \lambda}{s}\right)^2\right)} \ . \tag{3}$$

In Eq. 3, $\lambda$ corresponds to the location of the maximum and is equal to time 0 given that we chose to place the reaction barrier at the transition state. The scale parameter, $s$ is the second target to predict by using a GPR.

Rescaling $C_{ff}(t;T)$ by $C_{ff}(0;T)$ was found to be necessary to improve the prediction accuracy. While this requires future users to solve for Eq. 2 at time zero, it avoids the need to solve it at all subsequent times and hence reduces the cost to that of e.g. the Quantum Instanton approximation.[34]

We note that it was possible to use a Cauchy distribution here as there was no reflection or recrossing in the dynamics for any of the reactions. For other reactions, one would need to fit to a function which includes a negative contribution.

**Input Features**

Coulomb matrix (CM) and encoded bond (EB) input features were generated from reactant and product geometries with the MolML software package.[35] From these, a difference input feature was computed. Indeed difference input features have been successful in the context of machine learning kinetics.[16,36] Many other input feature representations could be considered, for instance graph input features or machine learned input features. While it is beyond the scope of this work, we plan on evaluating these in the future. To account for organic-metal atom interactions we considered the 6 atoms which moved the most after aligning reactant and product geometries. This number was chosen to ensure that at least one metal atom was included in the representation.

When training and testing GPR reaction rate constant product predictors, inverse temperature, $1/T$, and reaction energies (RE) were included as input features. The combined input features for the train and test set were scaled using the min-max (0,1) scaler fit to the train set implemented in the scikit-learn software package.[37] The target was not scaled or normalized. The same input features were used when training and testing GPRs to predict the Cauchy scale parameter (Eq 3).

**Search for optimal kernel and molecular input features for GPR fitting of the reaction rate constant products**

A search was carried out for the optimal geometric input features and GPR kernel combination. The optimal set was identified by comparing the train set mean absolute error (MAE) scores. We investigated both the use of single kernels (Matérn, radial basis function, pairwise with linear metric, rational quadratic and white noise) and sums of two single kernels.[38] For the temperature-wise split, it was found (Table 2 and SI Table S2) that Coulomb input features with min-max scaling together with a Matérn kernel summed to a pairwise linear kernel lead to the lowest MAE $9.92 \cdot 10^{-11}$ in units of the natural logarithm of inverse atomic units or time, log[1/au]. This corresponds to a MAPE of $2.41 \times 10^{-10}$ %. For the reaction-wise split, the optimal input features were the min-max scaled Coulomb input features together with 1/T and reaction energies and the optimal kernel was the sum of two Matérn kernels with a MAPE of $1.43 \times 10^{-9}$ % (Table 2, Table S3).

Table 2: Accuracy of optimal gaussian process regressors when predicting $\log(k(T) \cdot Q_R(T))$ or the Cauchy scale parameter, $s$, (Eq 3) for the fit of $C_{ff}(t)$. In the first column we list the optimal features and kernel. CM stands for Coulomb Matrix, 1/T for inverse temperature and RE for reaction energy. In the second column we specify whether the data was split based on temperature of reaction. The third column lists the target, $k(T)$ is the fully quantum reaction rate constant at a given temperature $T$, $Q_R(T)$ is the canonical reactant partition function and $C_{ff}(t)$ is the flux-flux correlation function at a given time $t$ and fixed temperature. In the fourth and fifth columns we report the mean absolute percent error (MAPE) on the train and test set.

| GPR Model | Data split | Target | *Fit* Train set MAPE | *Predict* Test set MAPE |
|---|---|---|---|---|
| *Input features:* Rescaled CM difference, 1/T & RE <br> *Kernel:* Matérn + Pairwise | Temperature-wise | $\log(k(T) \cdot Q_R(T))$ | $2.41 \times 10^{-10}$ | $5.02 \times 10^{-1}$ |
| *Input features:* Rescaled CM difference, 1/T & RE <br> *Kernel:* Matérn + Matérn | Reaction-wise | $\log(k(T) \cdot Q_R(T))$ | $1.43 \times 10^{-9}$ | $3.14 \times 10^{1}$ |
| *Input features:* Rescaled CM difference, 1/T & RE <br> *Kernel:* Matérn + Rational Quadratic | Temperature-wise | Scale parameter for Cauchy fit of $C_{ff}(t;T)/C_{ff}(0;T)$ | $1.32 \times 10^{-11}$ | $9.72 \times 10^{-1}$ |
| *Input features:* Rescaled CM difference, 1/T & RE <br> *Kernel:* Matérn + Rational Quadratic | Reaction-wise | Scale parameter for Cauchy fit of $C_{ff}(t;T)/C_{ff}(0;T)$ | $3.36 \times 10^{-11}$ | $2.94 \times 10^{1}$ |

**GPR prediction of reaction rate constant products**

The optimal GPR models listed in Table 2 were used to predict reaction rate constant products for previously unseen temperatures and reactions. For temperature-wise split test set values, the GPR had a high prediction accuracy with a MAPE of 0.5%. As we can see in Figure 3 the predicted values (y-axis) closely follow the exact values (x-axis). The highest accuracy was in the temperature range $T \in [250 - 350]K$ which is where there was a higher density of points in the training set (see SI Figure S4). On the other hand, in the case of reaction-wise split test values, the MAPE was significantly larger at 31.4%. In Figure 4 we compare the predicted and exact $k(T) \cdot Q_R(T)$ for the decomposition of activated CH$_3$* on a Pt(111) surface (reaction 8 of Table 1).

We find the Gaussian process regressor standard deviation to be large, however the predicted mean values are within one standard deviation from the exact values.

We believe these larger errors on reaction-wise split test data are due to the small size of the training dataset. To improve on this prediction error, in section "Computation of $k(T) \cdot Q_R(T)$, from a predicted fit of $C_{ff}(t;T)$", we look at training a GPR to predict a fit to the $C_{ff}(t;T)$ and integrating it to compute a $k(T) \cdot Q_R(T)$ value.

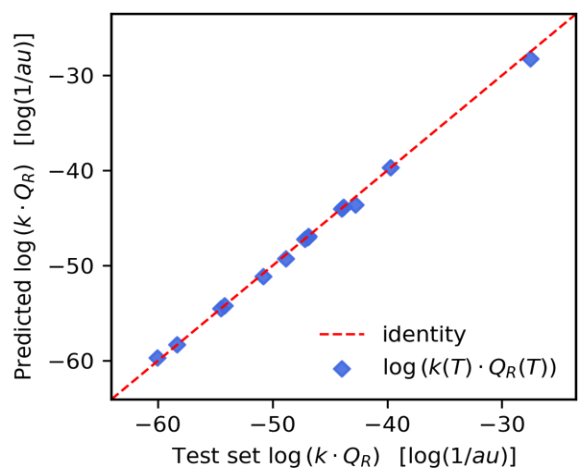

Figure 3: Pairwise plot of the predicted test set values of $\log(k(T) \cdot Q_R(T))$; the product of the quantum reaction rate constant $k(T)$ with the reactant canonical partition function $Q_R(T)$ for temperature-wise split train and test sets. The predicted values are in strong agreement with the test set with a test set MAPE of 0.502%.

**Search for optimal kernel for GPR fitting of scale parameter of Cauchy fit of flux-flux correlation function**

For the GPRs of Cauchy distribution $C_{ff}(t;T)/C_{ff}(t;0)$ fit scale parameters, $s$, (Eq 3), we used the same input features found optimal when fitting GPRs to predict $\log(k(T) \cdot Q_R(T))$; i.e. Coulomb matrix input features together with the inverse temperature and reaction energy. All input features were min-max scaled. We then searched for the optimal kernel and found that (Table 2, Table S4) for temperature-wise splitting the best was the sum of a Matérn kernel with a rational quadratic kernel. Here the training set MAPE was $1.32 \times 10^{-11}$%. For reaction-wise splitting the optimal kernel was a Matérn kernel summed to a rational quadratic kernel and the train set MAPE was $3.36 \times 10^{-11}$ % in atomic units of time.

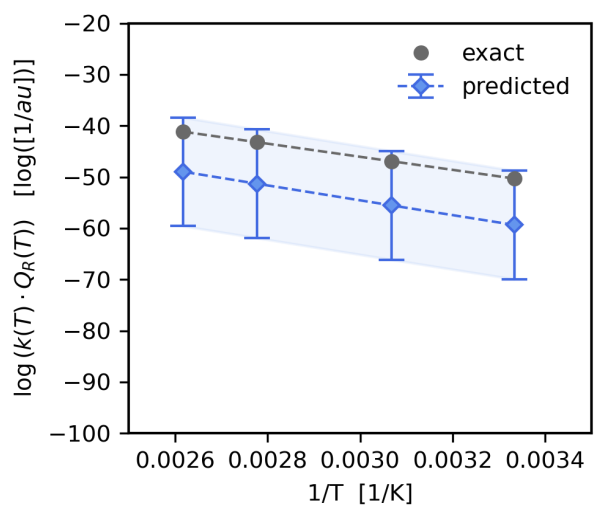

Figure 4: Predicted value of reaction rate constant product as a function of 1/T for the reaction of $CH_3^*+ ^* \rightarrow CH_2^* + H^*$ on Pt(111), taken from the test set (Table 1 – reaction 8). We see that the standard deviation of the trained GPR posterior predictor is quite large, however the predicted values are within one standard deviation from the exact results.

Table 3: The first two rows report the average mean absolute percent error on the computation of $C_{ff}(t;T)/C_{ff}(0;T)$ using a Cauchy distribution (Eq 3) with the GPR predicted scale parameter. The "rx" subscript indicates that MAE errors are averaged over all reactions. The last two rows report the percent mean absolute error on the logarithm of the product of the reaction rate constant with the reactant partition function. Here the product is obtained by using trapezium integration of the Cauchy fit of the scaled flux-flux correlation function.

| Computed quantity | Method | Error metric | % Error |
|---|---|---|---|
| $\dfrac{C_{ff}(t;T)}{C_{ff}(0;T)}$ | Cauchy curve fit with GPR predicted scale parameter for temperature-wise split data | $\langle MAPE\left(C_{ff}^{exact}(t;T), C_{ff}^{GPR\ Cauchy}(t;T)\right)\rangle_{rx}$ | $9.76 \times 10^{-1}$ |
| $\dfrac{C_{ff}(t;T)}{C_{ff}(0;T)}$ | Cauchy curve fit with GPR predicted scale parameter for reaction-wise split data | $\langle MAPE\left(C_{ff}^{exact}(t;T), C_{ff}^{GPR\ Cauchy}(t;T)\right)\rangle_{rx}$ | $9.95 \times 10^{-2}$ |
| $\log(k(T) \cdot Q_R(T))$ | Trapezium integral of predicted $C_{ff}(t)$ Cauchy fit for temperature-wise split data | $MAPE(\log(k \cdot Q_R)^{exact}, \log(k \cdot Q_R)^{GPR\ Cauchy})$ | $6.78 \times 10^{-1}$ |
| $\log(k(T) \cdot Q_R(T))$ | Trapezium integral of predicted $C_{ff}(t)$ Cauchy fit for reaction-wise split data | $MAPE(\log(k \cdot Q_R)^{exact}, \log(k \cdot Q_R)^{GPR\ Cauchy})$ | $8.21 \times 10^{-1}$ |

**GPR prediction of Cauchy fit flux-flux correlation functions**

With used the GPRs of the Cauchy fit scale parameter to obtain a "predicted" Cauchy fit of the scaled flux-flux correlation function. We found that these "predicted" fits closely followed the original fits of the exact scaled flux-flux correlation functions for both the temperature-wise split and reaction-wise split test data. In Figure 5 we show the pairwise correlation between the unscaled predicted fits (y-axis) and unscaled test set fits (x-axis) for the reaction wise split. The predictions are in strong agreement with the exact values. We note a series of segments in the predicted values of $C_{ff}(t;T)/C_{ff}(t;0)$. Each segment can be associated with one or at most two reactions. We also see accurate predictions of the flux-flux correlation function (Figure 7, panel a) for reaction 8 at 150K. The predicted values closely follow the exact values in time for this previously unseen test set reaction.

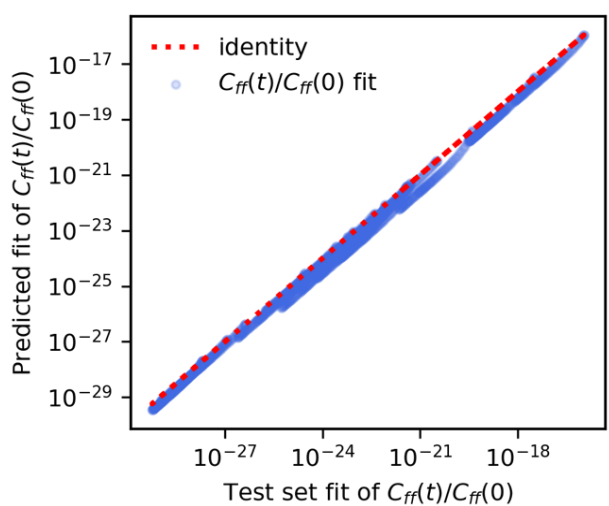

Figure 5: Plot of test set time series values from Cauchy fits of the flux-flux correlation function, normalized by its value at time zero (x-axis) respect to the time series values from predicted fits using GPRs (y-axis). Both axes are in log scale to emphasize data points. The predicted fits are in strong agreement with the exact Cauchy fits.

**Computation of $k(T) \cdot Q_R(T)$ from a predicted fit of $C_{ff}(t;T)$**

From our GPR Cauchy scale parameter predicted $C_{ff}(t;T)$ data we computed the reaction rate constant products by using trapezium integration for both the temperature wise and reaction wise split test set values. The results are shown in the last two rows of Table 3 and also in the pairwise plot (Figure 6). We note a very large improvement in the mean

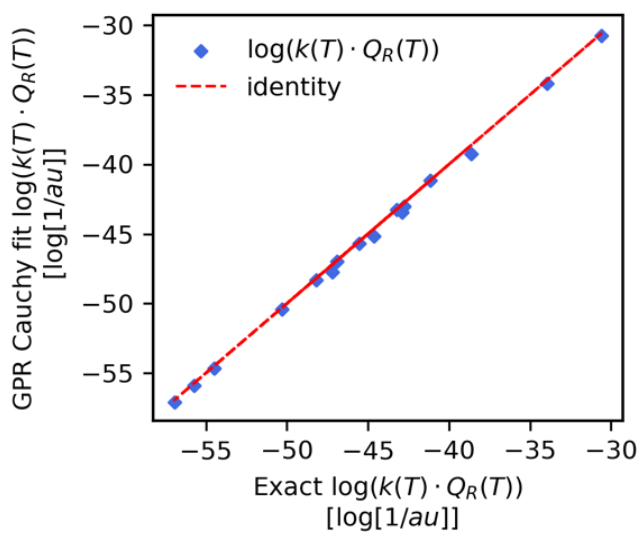

Figure 6: Pairwise plot of the exact (x-axis) respect to the integrated Cauchy GPR fit (y-axis) product of the reaction rate constant with the reactant partition function. Data was obtained for the reaction split previously unseen reaction. The MAPE = 0.82% is two orders of magnitude smaller than what was found when predicting the product directly for the same data (see Table 2)

absolute percent error (MAPE) for the reaction-wise split dataset. While the MAPE was on the order of $10^1$ when predicting the product directly (Table 2) it is now two orders of magnitude smaller and equal to 0.082%. If we look at the prediction of the reaction rate constant product for reaction 8 from the test set (Figure 7, panel b). we find a significant improvement respect to Figure 4. The predicted values now closely follow the exact as a function of temperature

With these last results we believe GPRs can predict kinetics not only as a function of temperature, but also for previously unseen reactions.

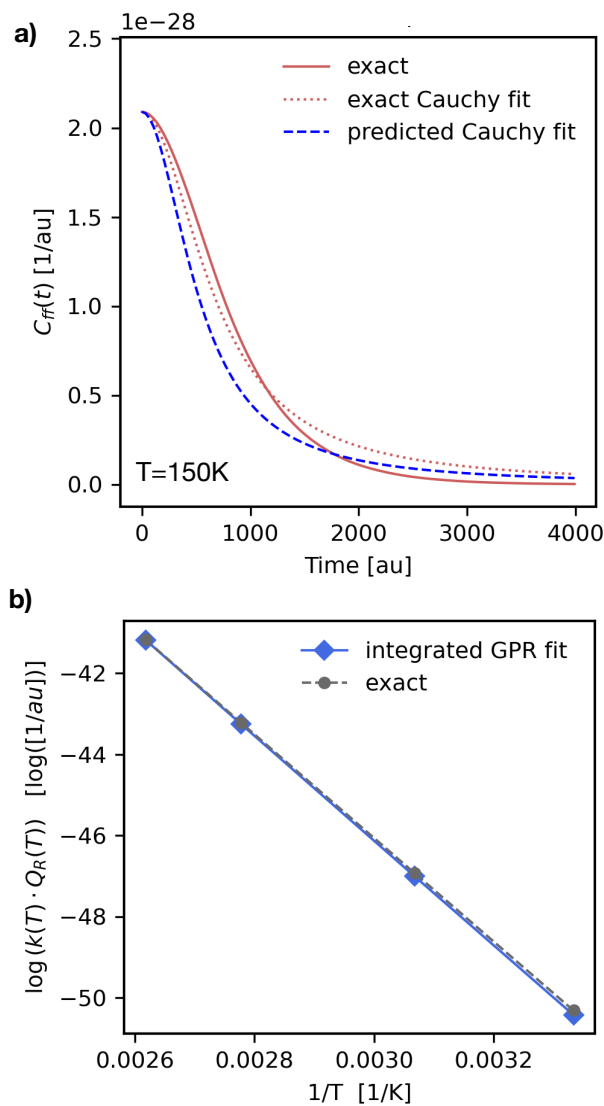

Figure 7: Panel a) Predicted Cauchy fit of the flux-flux correlation function (dashed blue line) for the reaction of $CH_3^* \rightarrow CH_2^* + H$ on Pt(111). (reaction 8 in Table 1 ) at 150K. Panel b) Comparison of the exact reaction rate constant products (grey circles and dashed line) with numerically integrated values obtained from the GPR Cauchy predicted scale parameter fit (blue diamonds and solid line). We see a large improvement on predicted test set reaction rate constant values respect to what we had found when predicting the rate constant product directly (Table 2).

**CONCLUSIONS**

Gaussian process regressors were trained on our home built set of exact flux-flux correlation functions and quantum reaction rate constant products for heterogeneous surface reactions.[33] We investigated the GPRs ability to predict both the product of the reaction rate constant with the reactant partition function and the flux-flux correlation function in time, rescaled respect to its initial value. Two previously unseen test sets were used to determine prediction ability, one contained previously seen reactions at new temperatures and the other previously unseen reactions at new temperatures. GPRs were accurate in predicting reaction rate constant products at new temperatures with a small MAPE on the order of $10^{-1}$. When looking at reaction-wise split however, the error was larger and on the order of $10^1$. We were able to reduce the prediction error by considering a different GPR target – the scale parameter of a Cauchy distribution function fit to the flux-flux correlation function. The resulting trapezium integrated reaction rate constants had an MAPE error which was two orders of magnitude smaller for that same reaction wise split test set 0.82%.

The cost of our predictors is low when looking at predicting the fully quantum $k(T) \cdot Q_R(T)$ directly. All that is needed in input is a reaction energy, temperature and the geometries of reactants and products. No information on minimum energy paths is required. When considering the prediction of the Cauchy fit for the flux-flux correlation function, one also needs the value of the flux-flux correlation function at time zero. While this comes with a cost it remains orders of magnitude smaller than the full computation of the flux-flux correlation function. We trust that this work will help in the prediction of fully quantum kinetic quantities.

## DATA AVAILABILITY

The flux-flux correlation function and reaction rate constant dataset can be found on Zenodo.[33]

## ACKNOWLEDGEMENTS

The authors would like to acknowledge Evan Komp for useful discussions on this work. The authors also acknowledge the Hyak supercomputer system at the University of Washington for support of this research.

# Supporting information for
# "Machine learning the quantum flux-flux correlation function for catalytic surface reactions"

Brenden G. Pelkie[*,1] and Stéphanie Valleau[*,2]

*Department of Chemical Engineering, University of Washington,
Seattle, Washington 98115, United States*

## TABLE OF CONTENTS

[1] email: bgpelkie@uw.edu
[2] email: valleau@uw.edu

## 1. MINIMUM ENERGY PATHS

Minimum energy paths were obtained from the Catalysis-hub energies of the reactants, products and transition states by fitting those values to single asymmetric Eckart barriers, $V_{\text{eckart}}(x)$ (Equation S1) or a function composed of the product of a logistic function and skewed normal function (Equation S2). The choice of the fitting function was determined based on the error of the fit.

$$V_{\text{eckart}}(x) = \frac{V_1(1-\alpha)}{1+e^{-\frac{2\pi x}{w_1}}} + \frac{V_1(1+\sqrt{\alpha})^2}{4\cosh^2\left(\frac{\pi x}{w_1}\right)} \tag{S1}$$

The Eckart barrier function depends on 3 parameters: $V_1$ is the height of the barrier, $w_1$ the width of the barrier, and $\alpha$ the asymmetry of the barrier.

$$V_{\text{skewnormal}}(x) = v * \underbrace{\left[\frac{L}{1+e^{-k(x-x_0)}}\right]}_{Logistic\ function} * \underbrace{\left[2*\frac{1}{\sigma\sqrt{2\pi}}e^{-\frac{1}{2}\left(\frac{x-\mu}{\sigma}\right)^2}\right] * \frac{1}{2}\left[1+\text{erf}\left(\frac{x*a-\mu}{\sigma\sqrt{2}}\right)\right]}_{Skewed\ normal\ distribution} \tag{S2}$$

The skewed normal barrier function depends on 5 parameters: $L$ the horizontal asymmetry of the barrier, $k$ the width of the logistic function, $a$ the horizontal asymmetry of the barrier, $v$ the overall vertical scaling of the barrier (used to fit activation energy), and $\sigma$ the overall width of the skewed normal distribution.

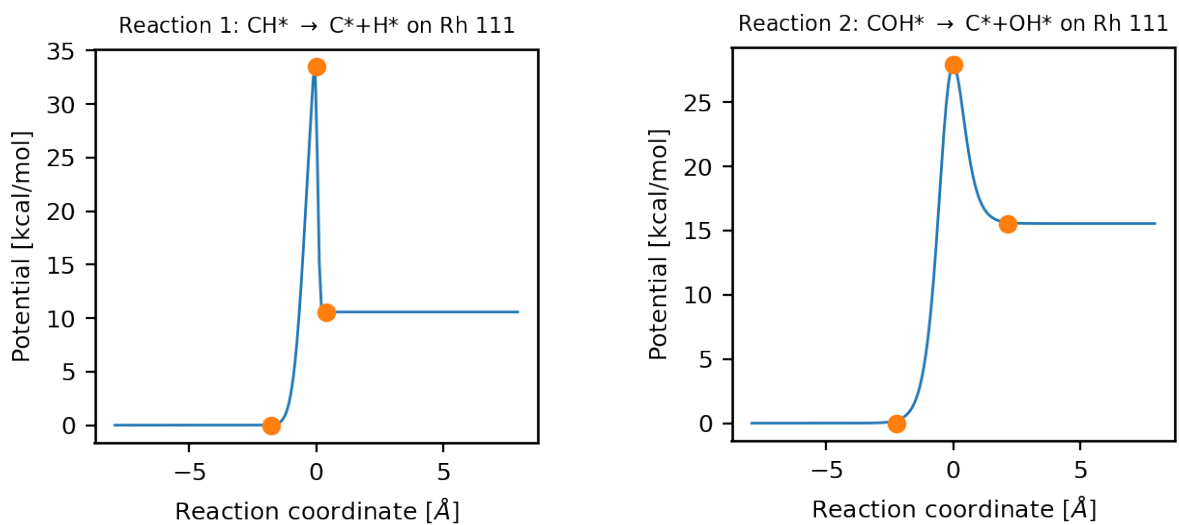

**Figure S1**: Plots of minimum energy path fits (skew logistic or Eckart) to reactions 1 and 2 reactant, product and transition state energies. See Table 1 of the manuscript for the list of reactions.

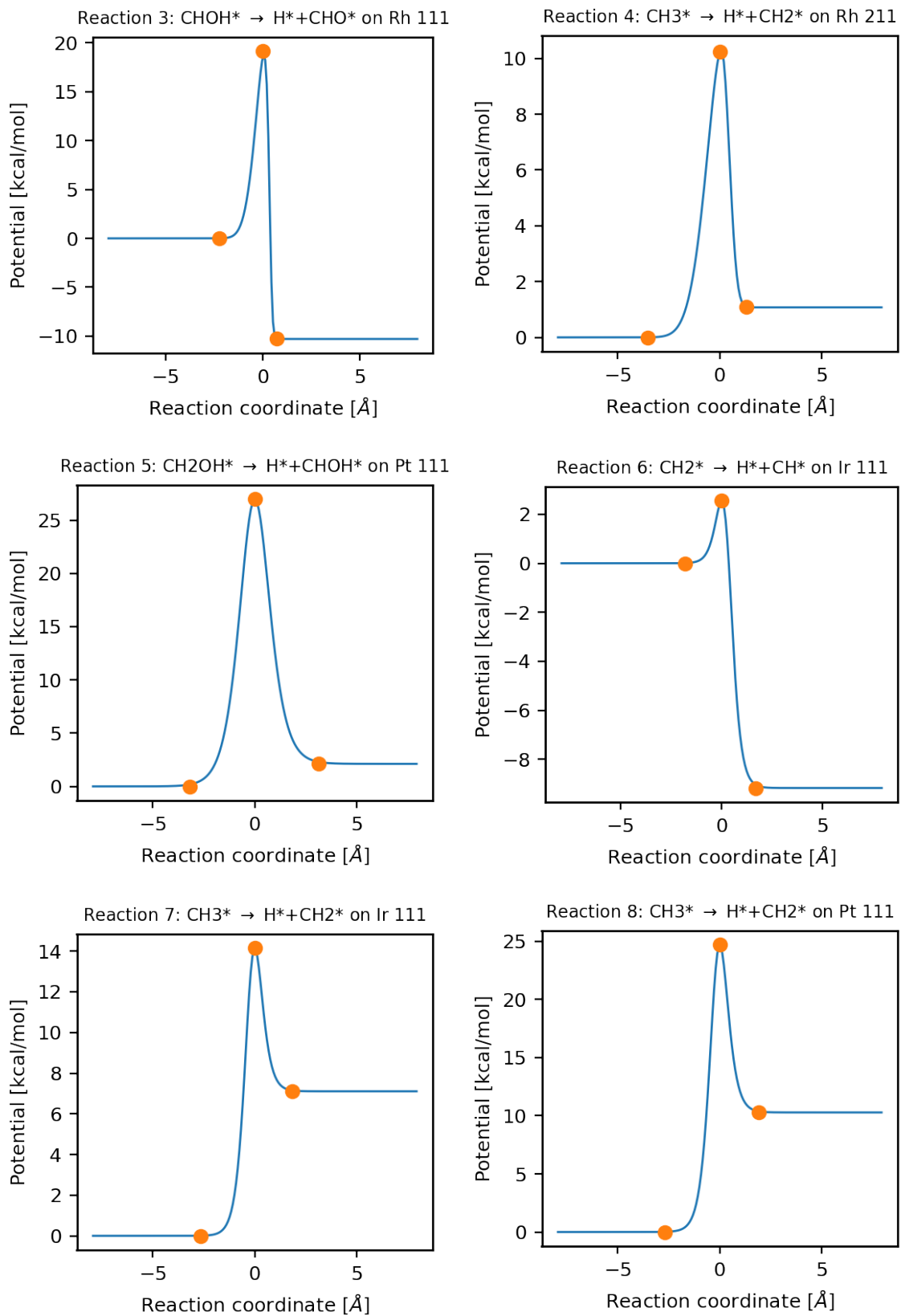

**Figure S2:** Plots of minimum energy path fits (skew logistic or Eckart) to reactions 3 through 8 reactant, product and transition state energies. See Table 1 of the manuscript for the list of reactions.

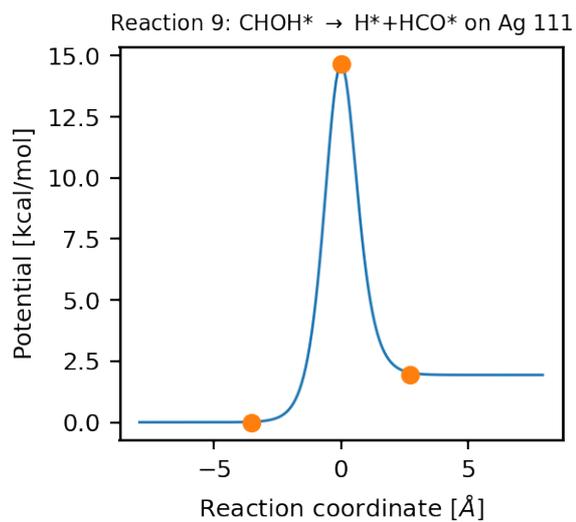
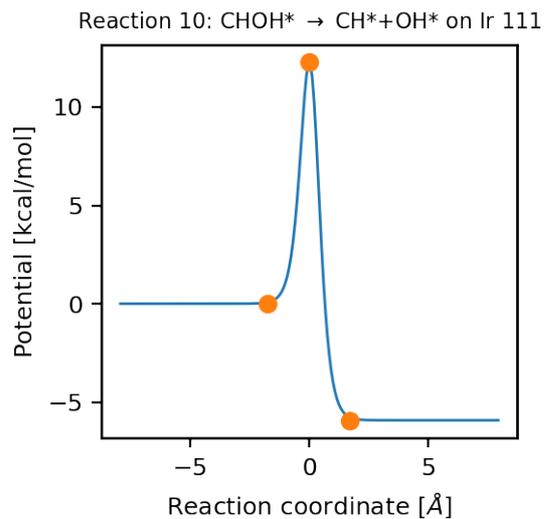
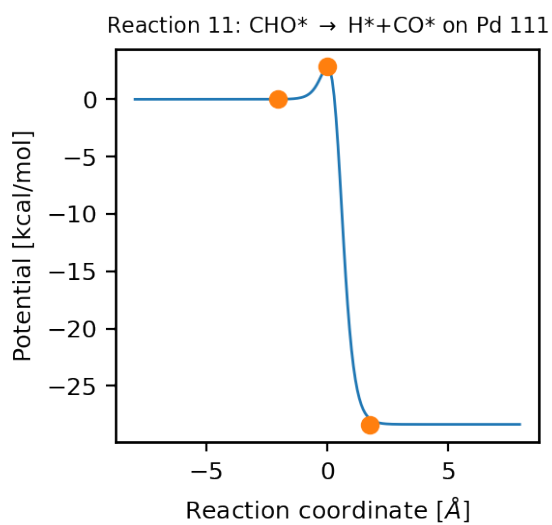
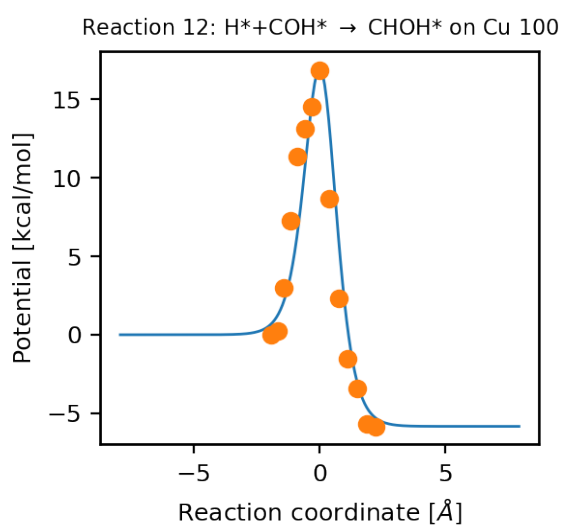
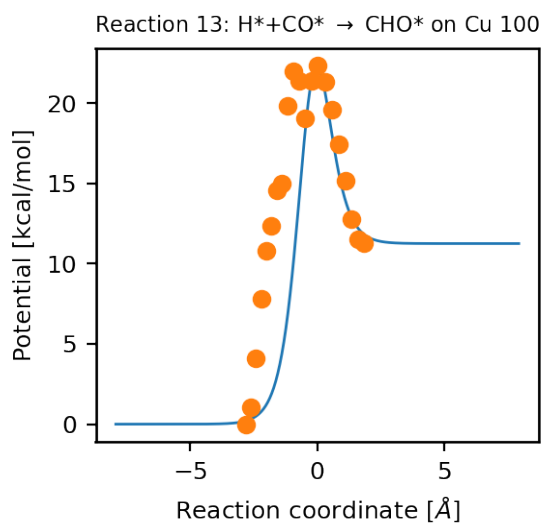
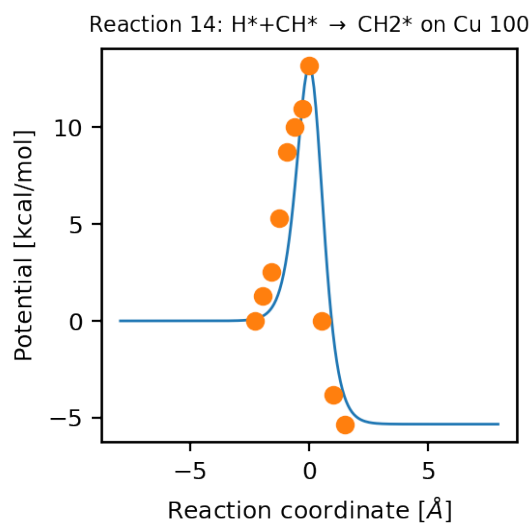

**Figure S3:** Plots of minimum energy path fits (skew logistic or Eckart) to reactions 9 through 14 reactant, product and transition state energies (reactions 9-11) or minimum energy paths (reactions 12, 13, and 14) (See Table 1 of the manuscript for a list of all reaction ).

## 2. DISCRETE VARIABLE REPRESENTATION AND $C_{ff}(t)$ CALCULATION

The parameters used for the DVR grid calculations are listed in Table S1.

**Table S1:** Description of the DVR grid parameters used to compute the flux-flux correlation function for each reaction. Randomly selected temperatures for each reaction are listed in the third column. The fourth column lists the spacing Δx in atomic units between points on the DVR grid. The fifth column lists the grid half-width, $L$. The complete grid covered the range [-$L$, $L$] with $N_{DVR}$ points.

| $N_{react}$ | Reaction | T [K] | Δx [au] | $L$ [au] | $N_{DVR}$ |
|---|---|---|---|---|---|
| 1 | $CH^* + ^* \to C^* + H^*$ | [300.0, 324, 356, 388] | 0.001 | 10 | 20001 |
| 2 | $COH^* + ^* \to C^* + OH^*$ | [300.0, 308, 356, 381] | 0.0075 | 8 | 2133 |
| 3 | $CHOH^* + ^* \to CHO^* + H^*$ | [300.0, 304, 355, 392] | 0.00075 | 10 | 26667 |
| 4 | $CH_3^* + ^* \to CH_2^* + H^*$ | [150.0, 180, 277, 394] | 0.0075 | 14 | 3733 |
| 5 | $CH_2OH^* + ^* \to CHOH^* + H^*$ | [300.0, 301, 359, 394] | 0.001 | 14 | 28001 |
| 6 | $CH_2^* + ^* \to CH^* + H^*$ | [150.0, 181, 310, 316] | 0.005 | 8 | 3201 |
| 7 | $CH_3^* + ^* \to CH_2^* + H^*$ | [150.0, 212, 284, 326] | 0.0075 | 8 | 2133 |
| 8 | $CH_3^* + ^* \to CH_2^* + H^*$ | [300.0, 326, 360, 382] | 0.01 | 8 | 1601 |
| 9 | $CHOH^* + ^* \to HCO^* + H^*$ | [200.0, 241, 276, 340] | 0.0075 | 14 | 3733 |
| 10 | $CHOH^* + ^* \to CH^* + OH^*$ | [200.0, 263, 278, 388] | 0.005 | 8 | 3201 |
| 11 | $CHO^* + ^* \to CO^* + H^*$ | [250.0, 254, 310, 373] | 0.005 | 8 | 3201 |
| 12 | $COH^* + H^* \to CHOH^*$ | [300.0, 321, 337, 383] | 0.0025 | 10 | 8001 |
| 13 | $CO^* + H^* \to CHO^*$ | [250.0, 296, 336, 381] | 0.0075 | 10 | 2667 |
| 14 | $CH^* + H^* \to CH_2^*$ | [250.0, 266, 321, 353] | 0.01 | 12 | 2401 |

Flux-flux correlation functions, $C_{ff}(t)$, were computed using software developed in-house on a time grid with a time spacing $\Delta t_1 = 35$ [au time] for the first 2000 au time, and $\Delta t_2 = 70$ [au time] for the rest of the time period. Calculations were run to a minimum stopping time of 10,000 [au time]. To calculate rate constant products $k(T) \cdot Q_R(T)$, the $C_{ff}(t)$ time series data was integrated using a trapezoidal integration scheme until a change of $C_{ff}(t)$ value of less than 1% was observed between 2 time points, or until time 8000 [au] was reached.

## 3. GRID SEARCH ON INPUT FEATURES FOR $k(T) \cdot Q_R(T)$

For every set of input features tested, the following GPR kernels, and every possible combination of 2 kernels, were iterated over:

- Matérn
- Radial Basis Function
- Rational Quadratic
- Pairwise Kernel
- White Noise Kernel

An initial kernel length scale of 1.0 and length scale bounds of (1x10-5, 1x10$^5$) were used for the Matérn and Radial basis function kernels. Gaussian Process Regressor training was done using 50 optimizer restarts.

**Table S2**: Representation and kernel hyperparameter grid search results for temperature split data. Green shading indicates selected input features and kernel. Optimal representations and kernels were selected based on train set MAE.

| Representation | Optimal kernel on train set | Train set MAE |
|---|---|---|
| Coulomb Matrix Difference, inverse temperature, and reaction energy, all minmax scaled | $60.8^2$ * Matern(length_scale=4.49, nu=1.5) + $15.5^2$ * PairwiseKernel(gamma=1997.944185658677, metric=linear) | 9.92x10$^{-11}$ |
| Encoded Bonds Difference, inverse temperature, and reaction energy, all minmax scaled. | $7.69^2$ * Matern(length_scale=29.6, nu=1.5) + $306^2$ * Matern(length_scale=29.6, nu=1.5) | 1.11x10$^{-10}$ |

**Table S3:** Representation and kernel hyperparameter grid search results for reaction split. Green shading indicates selected input features and kernel. Optimal representations and kernels were selected based on train set MAE.

| Representation | Optimal kernel on train set | Train set MAE |
|---|---|---|
| Coulomb Matrix Difference, inverse temperature, and reaction energy, all rescaled (minmax scaler) | $128^2$ * Matern(length_scale=11.5, nu=1.5) + $0.0114^2$ * Matern(length_scale=189, nu=1.5) | 5.69x10$^{-10}$ |
| Encoded Bonds Difference, inverse temperature, and reaction energy, all rescaled (minmax scaler) | $7.31^2$ * PairwiseKernel(gamma=0.004392928734334214, metric=linear) + $316^2$ * Matern(length_scale=35.3, nu=1.5) | 6.60x10$^{-10}$ |

## 4. INPUT FEATURES FOR $C_{ff}(t)$ CAUCHY CURVE FIT

Coulomb matrix difference geometry features were used for the $C_{ff}(t)$ Cauchy curve fit problem, as these were found to work best for the $kQ(T)$ problem. Coulomb matrix features, inverse temperature, and reaction energy were used as input features and were rescaled using the minmax scaler. The Cauchy fit scale parameter was also rescaled.

**Table S4:** Best kernels based on train set MAE for Cauchy curve fit parameter prediction task

| Split | Fit kernel parameters | Train set MAE |
|---|---|---|
| Temperature | $316^2$ * Matern(length_scale=1e+05, nu=1.5) + $316^2$ * RationalQuadratic(alpha=1.02, length_scale=1.35) | $5.86 \times 10^{-11}$ |
| Reaction | $223^2$ * Matern(length_scale=2.15, nu=1.5) + $141^2$ * PairwiseKernel(gamma=0.000679, metric=linear) | $6.71 \times 10^{-11}$ |

## 5. PREDICTED $k(T) \cdot Q_R(T)$ USING TEMPERATURE SPLIT

Here we report the role of undersampled temperature values on the $k(T) \cdot Q_R(T)$ prediction accuracy as a function of temperature (Figure S4).

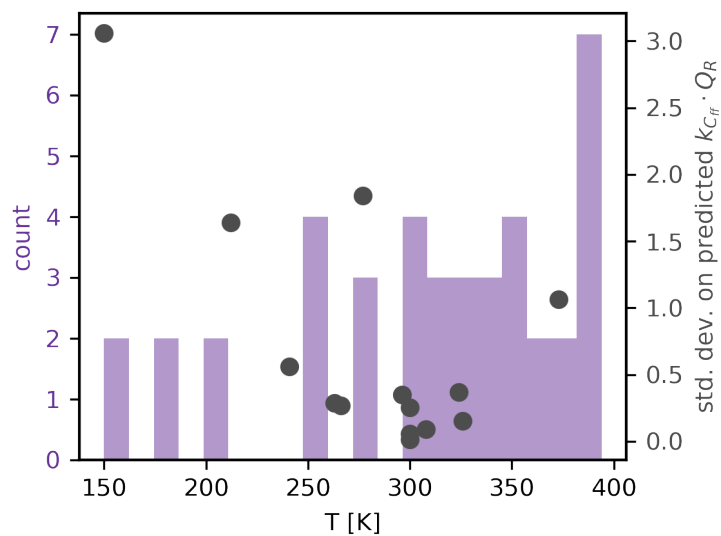

**Figure S4:** (y-axis left hand side) count of entries in training dataset binned by temperature (x-axis) compared to (y-axis right hand side) standard deviation on predicted value of $k(T) \cdot Q_R(T)$ for test set entries. We see that where the density of training points is higher the predicted error is lower, as expected.